\documentclass[a4paper,11pt]{article}

\usepackage{filecontents}

\usepackage[utf8]{inputenc}
\usepackage[T1]{fontenc}
\usepackage{lmodern}
\usepackage{geometry}
\usepackage{graphicx}
\usepackage{authblk}
\usepackage{url}
\usepackage{cite}
\usepackage{fancyhdr}
\usepackage[hidelinks]{hyperref}

\usepackage{amsmath}
\usepackage{orcidlink}

\usepackage{float}
\usepackage{amssymb}
\usepackage{doi}
\usepackage{lineno}
\usepackage{etoolbox}

\AtBeginEnvironment{figure}{}
\AtEndEnvironment{figure}{}
\AtBeginEnvironment{table}{}
\AtEndEnvironment{table}{}
\makeatother

\fancypagestyle{firstpage}{
	\fancyhf{}
	\fancyhead[C]{\includegraphics[height=2.2cm]{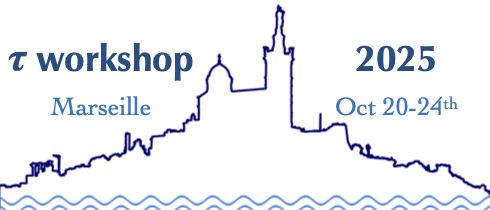}\\ \textit{Proceedings of the 18th International Workshop on Tau Lepton Physics}
	} 

}

\title{Wa$\nu$ePID: Studies of DOM-level waveform timing for track vs. cascade discrimination in IceCube at 5-100 GeV}
\author{Steven Young Eulig}

\affil{Department of Physics and Laboratory for Particle Physics and Cosmology (LPPC), Harvard University, Cambridge, MA, 02138, USA}

\date{}

\begin{document}

\maketitle
\thispagestyle{firstpage}

\vspace{-1cm}
\begin{center}
	\textit{On behalf of the IceCube Collaboration}
\end{center}
\vspace{0.5cm}


\begin{abstract}
The IceCube Neutrino Observatory is a cubic-kilometer Cherenkov detector embedded in the Antarctic ice at the South Pole. Its densely instrumented sub-array and dedicated low-energy analyses provide sensitivity to neutrinos in the 5–100~GeV range, enabling precision studies of neutrino oscillations and searches for new physics. This work focuses specifically on this low-energy regime, where sparse hit patterns limit the performance of topology-based reconstruction and classification methods. We introduce Waveform-based Particle Identification (\textsc{WavePID}), a statistically rigorous and interpretable likelihood-ratio discriminator for track--cascade separation, built from Monte Carlo templates in timing-aware, physics-motivated observables and validated through dedicated simulations. Applied to both Monte Carlo and 11.1~years of IceCube data, \textsc{WavePID} suggests improved cascade purity by about 5 percentage points at a fixed 20\% down-selection rate relative to the current leading cascade selection, while maintaining Data--MC agreement within detector systematics. The approach is compact and robust to sparse observations, demonstrating the value of waveform-level timing for low-energy reconstruction.
\end{abstract}

\enlargethispage{-1cm}

\section{Introduction}
The IceCube Neutrino Observatory is a cubic-kilometer array of optical sensors deployed 1.5--2.5~km deep in the Antarctic ice. A part of it is DeepCore, a more densely instrumented subarray optimized for events in the few- to hundred-GeV range~\cite{IceCube_DeepCore_2012,Aartsen_2017}. It offers sensitivity to neutrino oscillation phenomena, such as the determination of atmospheric parameters, the neutrino mass ordering, and searches for new physics including sterile neutrinos~\cite{IceCube_mixing_2023,IceCube_mass_ordering_2020,IceCube_steriles_2024,cabrera2025searchingsubevsterileneutrinos}. Neutrino interactions in or near the instrumented volume produce charged secondaries that emit Cherenkov photons, which are detected by Digital Optical Modules (DOMs). Each DOM employs an Analog Transient Waveform Digitizer (ATWD) to capture nanosecond-scale structure of the electrical signal. The resulting digitized signals are processed into calibrated pulse time series with charges and timestamps used for downstream reconstruction and analysis.

Subsequent reconstruction algorithms classify events as track-like or cascade-like based on the geometry and timing of the observed light. Track-like events, typically arising from charged-current \(\nu_\mu\)-CC interactions, produce elongated light patterns that trace muon trajectories. The produced muons are minimally ionizing particles at these energies and can travel hundreds of meters in ice. Cascade-like events, associated with charged-current \(\nu_e\) and \(\nu_\tau\) interactions or neutral-current interactions of all flavors, produce compact energy depositions that are intrinsically elongated over a few meters but yield an approximately isotropic light pattern after photon scattering in the ice. Accurate separation of these topologies is important for flavor-dependent analyses, oscillation measurements, and background suppression.

The task of separating tracks from cascades presents a significant challenge within the $5$–$100\text{ GeV}$ energy range in IceCube. This difficulty arises because the reduced particle energy in this regime results in diminished Cherenkov light yields and shorter particle tracks, severely limiting the signal information available for discrimination. With many events triggering fewer than $20$ Digital Optical Modules (DOMs), the resulting limited event topology significantly impairs standard reconstruction algorithms. This low DOM multiplicity and lack of spatial differentiation necessitate the utilization of waveform-level timing information recorded by each DOM to achieve robust particle identification. The use of waveform-level timing has previously been explored in the Fermilab water-Cherenkov experiment FNAL-1267 at 2–8~GeV, where statistical separation between minimally ionizing (track-like) and non-minimally ionizing (cascade-like) events was achieved purely from waveform information~\cite{Samani_2020}.
Here, in this work, we introduce \textsc{WavePID}, a likelihood-based fit for track vs.~cascade discrimination and investigate first-order effects of the underlying microphysics with \textsc{Geant4} simulations (a software package that simulates particle interactions within matter).

\section{IceCube study and test statistic evaluation}
\begin{figure}
    \centering
    \includegraphics[width=0.65\linewidth]{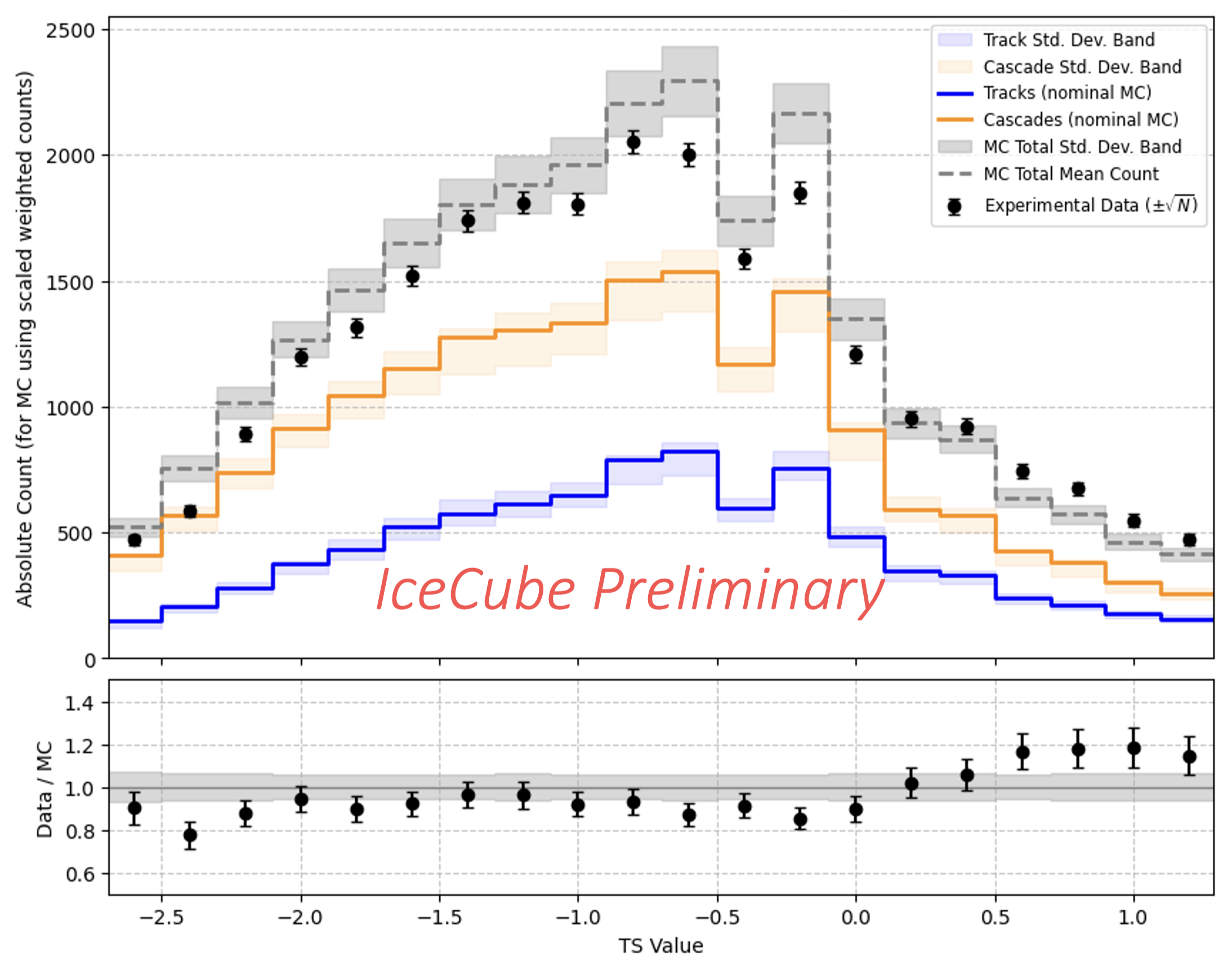}
    \caption{Distribution of the \textsc{WavePID} TS for true tracks and cascades. Shaded bands indicate one-standard-deviation detector systematics. Data are overlaid for comparison.}

    \label{fig:tsdist}
\end{figure}
\begin{figure}[H]
    \centering
    \includegraphics[width=0.5\linewidth]{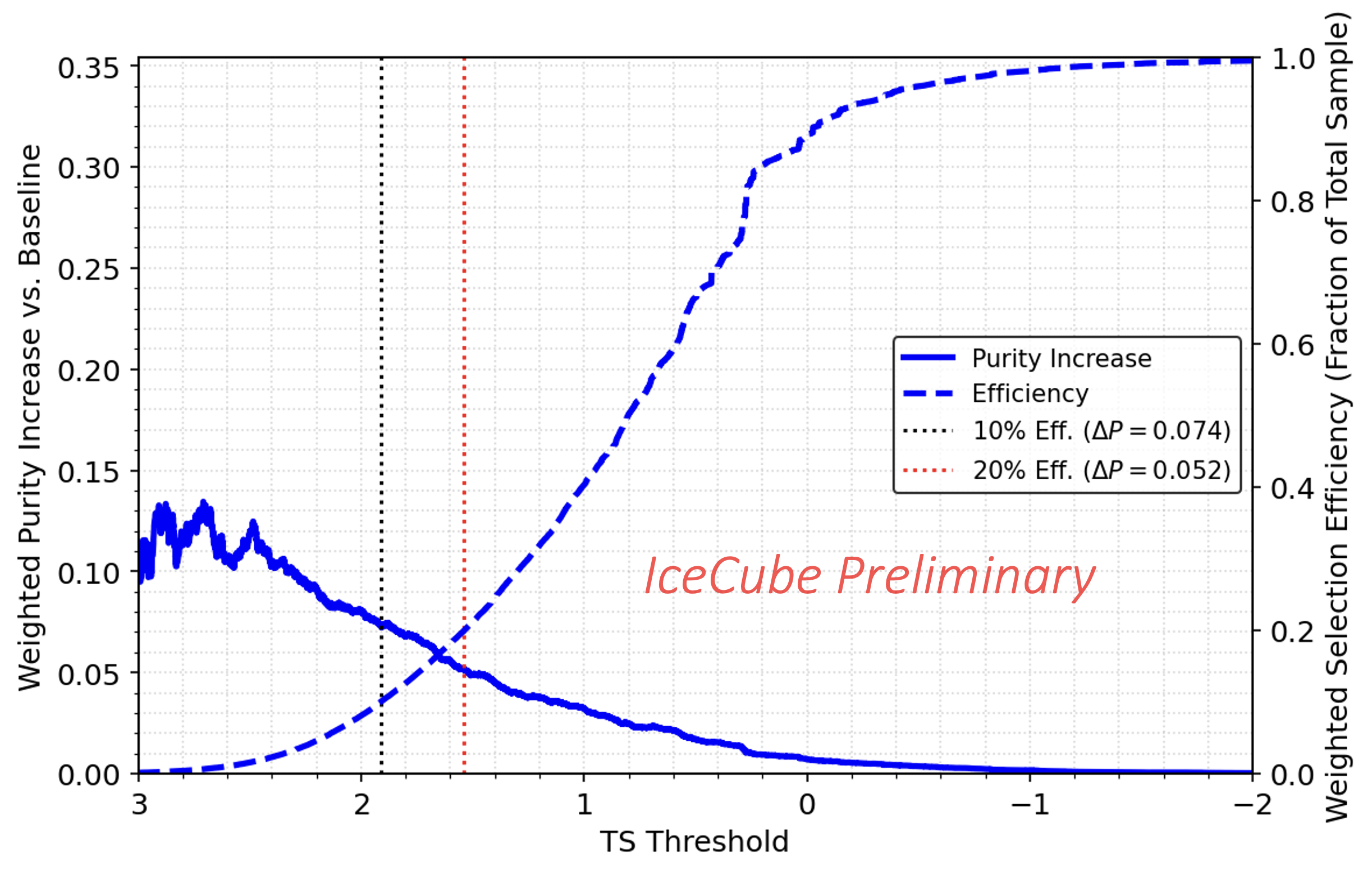}
\caption{Cascade purity gain and selection efficiency from the \textsc{WavePID} TS within the \textsc{DynEdge} cascade bin for events with $10 \leq N_{\text{hits}} < 20$~hit DOMs (on the nominal MC set).}

    \label{fig:boost}
\end{figure}

Given the sparse light deposition typical of IceCube events at 5-100 GeV energy range, \textsc{WavePID} is used to investigate its reliability in distinguishing track from cascade topologies. The method is a template-based test statistic (TS) built from two observables: the fraction of charge recorded in the first few nanoseconds and the distance between each DOM and the reconstructed event vertex (see \nameref{sec:appendix} for details). These observables capture differences in waveform shapes and account for the effects of photon propagation in the ice. Figure~\ref{fig:tsdist} shows the TS distribution for true tracks and cascades, including shaded bands that represent one standard deviation across detector systematics, compared to data. The current state-of-the-art low-energy particle identification (PID) approach in IceCube, \textsc{DynEdge}~\cite{IceCube_DynEdge_2022}, exhibits reduced performance in the cascade-PID bin, especially for events with $\leq 20$~hit DOMs. In this bin, where the cascade purity is about 66\% (under a set of standard cuts that were also used in this work), the proposed \textsc{WavePID} TS enhances cascade purity by approximately 5\% while retaining 20\% of the selected sample, as shown in Figure~\ref{fig:boost}.

\section{Photon arrival time distributions in Geant4}
\begin{figure}[H]
    \centering
    \includegraphics[width=1\linewidth]{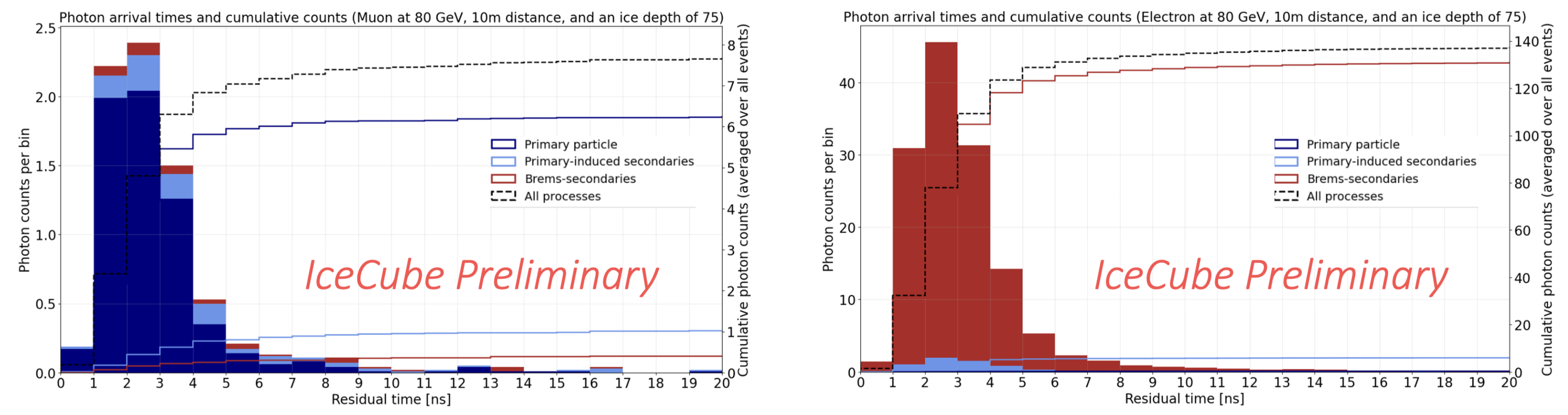}
    \caption{PATDs for \(\mu\) (left) and \(e\) (right) resolved by parent process.}

    \label{fig:PATDs}
\end{figure}

In this section, we examine the data-generating process to identify waveform features that distinguish tracks from cascades. Using dedicated simulations, we compute photon arrival time distributions (PATDs) for a single-DOM configuration embedded in ice. The simulation is built in C++ using \textsc{Geant4}~\cite{GEANT4:2002zbu} and the IceCube OMSim framework~\cite{OMSim}, which we extend here to trace the origin of each photon back to the generating process. The study simulates electrons and muons with energies between 5 and 100~GeV, DOM--vertex distances from 5 to 50~m, and incidence angles covering the full solid angle. The resulting PATDs (Figure~\ref{fig:PATDs}) show that minimally ionizing muons produce a prompt peak dominated by Cherenkov photons from the primary muon, whereas electron-induced cascades generate a broader distribution with light originating mainly from Bremsstrahlung secondaries. These photon-level differences are reflected in the waveform timing behavior discussed in Sec.~2 and motivate the use of the early-charge fraction as an input to the template-based test statistic employed by \textsc{WavePID}. The quantitative Geant4 results are only used to examine first-order photon-level physics. It does not include PMT response, DAQ electronics, waveform calibration, or the spatial and directional variability of the in-ice optical model. All WavePID templates, performance estimates, and systematics bands are derived from the full IceCube simulation and 11.1 years of data, outlined in Section~2.

\begin{figure}[H]
    \centering
    \includegraphics[width=0.45\linewidth]{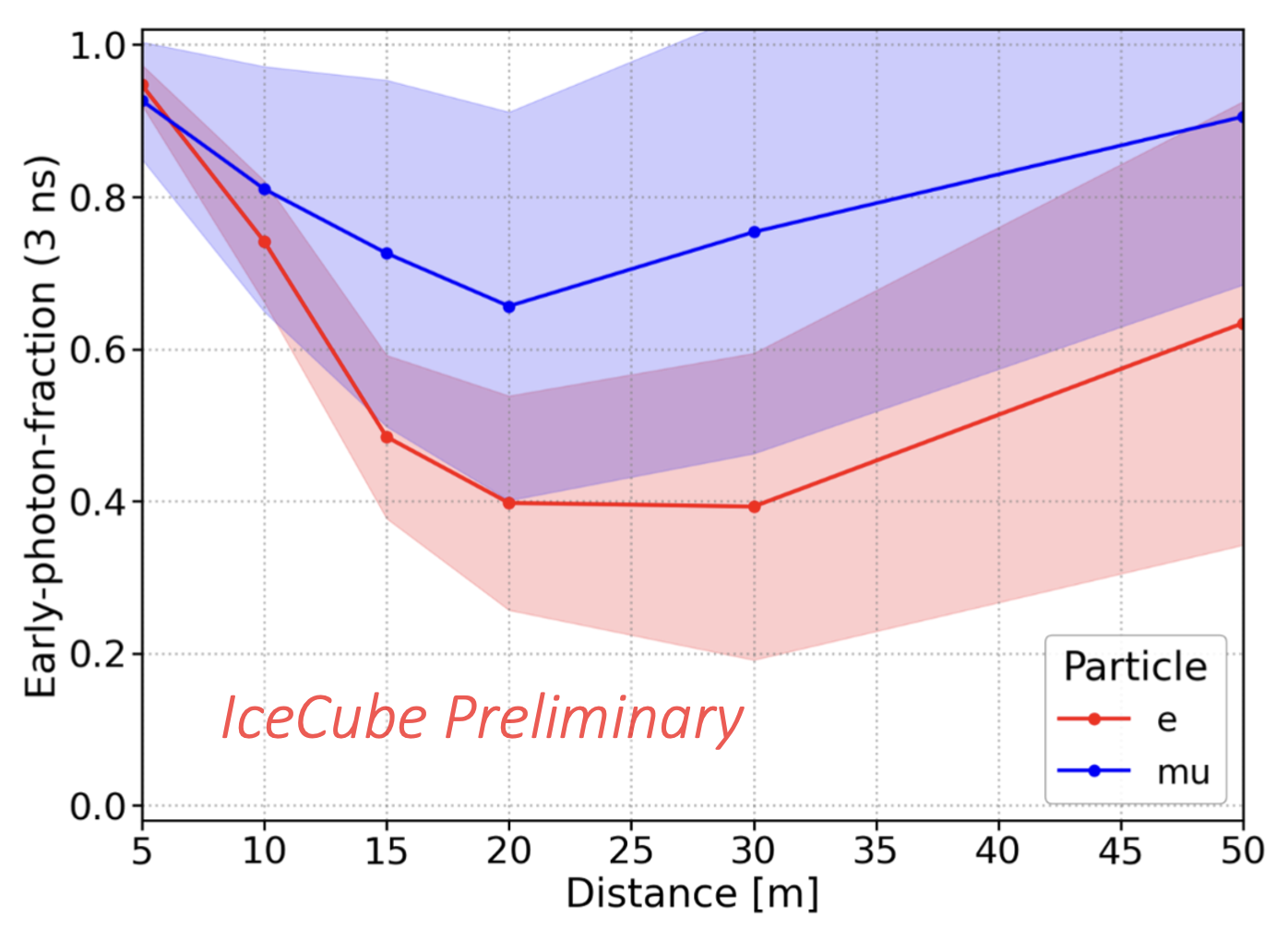}
    \caption{Early-photon fraction at \(80~\mathrm{GeV}\) as a function of distance. Shaded bands indicate \(1\sigma\) variations.}

    \label{fig:g4curve}
\end{figure}

A practical way to characterize the PATD shape is through the early-photon fraction, defined as the proportion of all detected photons that reach the DOM within a selected time window. Figure~\ref{fig:g4curve} shows the early-photon fraction within a 3~ns window as a function of the distance between the DOM and the injected charged particle. Muons exhibit consistently higher early-photon fractions than electrons across all distances that were investigated here, reflecting the prompt Cherenkov light from the primary muon track. Electrons produce lower values due to the more diffuse emission from secondary Bremsstrahlung in the cascade. The difference is most pronounced for particle injections about 20--30~m above the DOM. This behavior motivates the inclusion of the DOM--vertex distance as an input to the \textsc{WavePID} TS. The joint dependence on distance and energy indicates that these features are inherently multivariate, which supports the use of a likelihood-ratio TS rather than a fixed early-charge-fraction threshold or width of the waveform's primary peak. Simulations further show that DOM rotations in zenith or azimuth have negligible impact on the PATD shape, affecting only the total light yield.

\section{Conclusion and outlook}
To improve cascade selection below 100~GeV in IceCube, particle-identification methods must handle sparse hit patterns. This work introduces a template-based log-likelihood-ratio test statistic that increases cascade purity at 5--100~GeV and low hit multiplicities. Data and Monte Carlo simulations show agreement within statistical and detector uncertainties. The approach is transparent and physically interpretable, relying on two variables directly motivated by photon timing and geometry. This is the first work demonstrating that water-Cherenkov measurements conducted at Fermilab \cite{Samani_2020} are directly applicable to PID in neutrino telescopes like IceCube. The resulting \textsc{WavePID} performance can be further evaluated for its potential to enhance the sensitivity of IceCube's oscillation analyses.

The observed interplay between photon timing, the number of pulses, and the DOM--vertex distance suggests that more advanced particle-identification methods could achieve additional gains. The current \textsc{DynEdge} method is already based on Graph Neural Networks and future studies may employ other machine learning techniques such as transformer-based variational autoencoders to learn efficient event representations~\cite{yu2024learningefficientrepresentationsneutrino}. This work motivates future studies aimed at distinguishing electromagnetic from hadronic cascades, a capability that will be significantly enhanced by the IceCube Upgrade. The Upgrade's planned denser instrumentation and incorporation of multi-PMT DOMs are expected to boost cascade statistics and greatly improve the reconstruction of low-energy events~\cite{IceCube_Gen2_2021}. Next, we will explore global event-level timing as an additional test-statistic variable, since minimally ionizing tracks propagate faster than the Cherenkov photons they produce in ice.

\newpage

\bibliographystyle{unsrt}
\bibliography{references}

\newpage
\section{Appendix}\label{sec:appendix}

For the test statistic, we model the binned hits of event \(e\) under class \(s\in\{t,c\}\) as independent Poisson counts
with means \(\mu^{(e)}_{s,ij}=\alpha_{e,s}\,\hat{\lambda}_s(i,j)\), where \(\alpha_{e,s}>0\) is a per–event rate and \(\hat{\lambda}_s\) encodes only shape with \(\sum_{i,j}\hat{\lambda}_s(i,j)=1\).
The likelihood is
\[
\mathcal{L}_s(e)
=
\prod_{i,j}
\mathrm{Pois}\!\bigl(H_e(i,j)\mid \mu^{(e)}_{s,ij}\bigr)
=
\prod_{i,j}
\frac{e^{-\mu^{(e)}_{s,ij}}\bigl(\mu^{(e)}_{s,ij}\bigr)^{H_e(i,j)}}{H_e(i,j)!}\, .
\]
The log–likelihood ratio between cascade and track hypotheses is
\[
\log\frac{\mathcal{L}_c(e)}{\mathcal{L}_t(e)}
=
\sum_{i,j} H_e(i,j)\,
\log\!\frac{\hat{\lambda}_c(i,j)}{\hat{\lambda}_t(i,j)}
\;+\; N_e\log\!\frac{\alpha_{e,c}}{\alpha_{e,t}}
\;-\; \bigl(\alpha_{e,c}-\alpha_{e,t}\bigr),
\]
where \(N_e=\sum_{i,j}H_e(i,j)\). Maximizing each likelihood over \(\alpha_{e,s}\) and under \(\sum_{i,j}\hat{\lambda}_s=1\) yields \(\hat{\alpha}_{e,s}=N_e\), so the rate terms cancel:
\[
N_e\log\!\frac{\hat{\alpha}_{e,c}}{\hat{\alpha}_{e,t}} - (\hat{\alpha}_{e,c}-\hat{\alpha}_{e,t})
= N_e\log 1 - (N_e-N_e)=0.
\]
Therefore the WavePID test statistic reduces to the log–likelihood ratio of shapes,
\[
\mathrm{TS}(e)
=
\log\frac{\mathcal{L}_t(e)}{\mathcal{L}_c(e)}
=
\sum_{i,j} H_e(i,j)\,
\log\!\frac{\hat{\lambda}_t(i,j)}{\hat{\lambda}_c(i,j)}
=
\sum_{n\in e}\log\!\frac{\hat{\lambda}_t(b_n)}{\hat{\lambda}_c(b_n)},
\]
where \(b_n\) denotes the bin containing \(\mathbf{x}_n^{(e)}\). An event with a high TS-value therefore has a higher probability of being a track event. The following bin edges were selected empirically:
\begin{equation*}
\begin{aligned}
\mathbf{e}_r &= (0, 10, 20, 30, 40, 50, 60, 75, 100, 200, 500)\ \mathrm{m},\\
\mathbf{e}_\xi &= (0.0, 0.1, 0.2, \ldots, 0.9, 1.0).
\end{aligned}
\end{equation*}

The dataset is filtered to events with ($5-100~\mathrm{GeV}$), ($-1\leq \mathrm{cos}(\theta) \leq 0.3$), DeepCore containment, and a high neutrino probability (l7\_muonclassifier\_probnu$>0.8$) and to hits with an ATWD-flag and at least 2 distinct reconstructed pulses. The systematics are evaluated by computing the standard deviation across 31 different MC sets, which represent the IceCube detector uncertainties (parametrized via 6 parameters for DOM efficiency, hole ice, bulk ice absorption and scattering, and dust layer ice absorption).

\end{document}